\documentclass[preprint,showpacs,showkeys] {revtex4}
\usepackage{amssymb}
\usepackage{amsmath}
\usepackage{graphicx}

\begin{document}

\title{Long-Range Order and Dynamic Structure Factor of a Nematic
under a Thermal Gradient}

\author{R. F. Rodr\'{\i}guez}
\altaffiliation[Also at ]{FENOMEC. Fellow of SNI, Mexico.}
\email{zepeda@fisica.unam.mx}

\author{H. H\'{\i}jar}

\affiliation{Instituto de F\'{\i}sica. Universidad Nacional
Aut\'{o}noma de M\'{e}xico.\\
Apdo. Postal 20-364, 01000 M\'{e}xico, D. F., M\'{e}xico.}

\date{21th September 2005}

\begin{abstract}
We use a fluctuating hydrodynamic approach to calculate the orientation
fluctuations correlation functions of a thermotropic nematic liquid crystal
in a nonequilibrium state induced by a stationay heat flux. Since in this
nonequilibrium stationary state the hydrodynamic fluctuations evolve on
three widely separated times scales, we use a time-scale perturbation
procedure in order to partially diagonalize the hydrodynamic matrix. The
wave number and frequency dependence of these orientation correlation
functions is evaluated and their explicit functional form on position is
also calculated analytically in and out of equilibrium. We show that for
both states these correlactions are long-ranged. This result shows that
indeed, even in equilibrium there is long-range orientational order in the
nematic, consistently with the well known properties of these systems.We
also calculate the dynamic structure of the fluid in both states for a
geometry consistent with light scattering experiments experiments. We find
that as with isotropic simple fluids, the external temperature gradient
introduces an asymmetry in the spectrum shifting its maximum by an amount
proportional to the magnitude of the gradient. This effect may be of the
order of 7 per cent. Also, the width at half height may decrease by a factor
of about 10 per cent. Since to our knowledge there are no experimental results
available in the literature to compare with, the predictions of our model
calculation remains to be assessed.
\end{abstract}



\maketitle

\section{Introduction}

Thermal fluctuations in an equilibrium isotropic fluid always give rise to
short-range equal-time correlation functions, except close to a critical
point. However, when external gradients are applied, equal-time correlation
functions may develop long-range contributions, whose nature is very
different from those in equilibrium. For a variety of systems in
nonequilibrium states it has been shown theoretically that the existence of
the so called generic scale invariance is the origin of the long range
nature of the correlation functions, \cite{grinstein1}, \cite{ronis3}. For
instance, for a simple fluid under a thermal gradient its structure factor,
which determines the intensity of the Rayleigh scattering, diverges as $%
q^{-4}$ for small values of the wave number $q$. This dependence amounts to
an algebraic decay of the density-density correlation function, a feature
that has been verified experimentally \cite{beysens}, \cite{law1}.

Although studies of the behavior of fluctuations in nematic liquid crystals
about nonequilibrium states are more scarce, some specific examples have
been studied theoretically. Such is the case of the nonequilibrium
situations generated by a static temperature gradient \cite{brand1}, a
stationary shear flow \cite{brand2} or by an externally imposed constant
pressure gradient \cite{rodriguez1}, \cite{rodriguez2}, \cite{camacho}. In
the first two cases it was found that the nonequilibrium contributions to
the corresponding light scattering spectrum were small, but in the case of a
Poiseuille flow induced by an external pressure gradient the effect may be
quite large. To our knowledge, however, at present there is no experimental
confirmation of these effects, in spite of the fact that for nematics the
scattered intensity is several orders of magnitude larger than for ordinary
simple fluids \cite{handbook}.

In this work we present a model calculation based on a fluctuating
hydrodynamic description with a time-scale perturbation formalism to
calculate analytically the equal-time correlation functions of \ the
transverse and longitudinal orientation components of a thermotropic nematic
liquid crystal. We derive the explicitly space dependence of these
quantities and show that they exhibit long-range order not only in
equilibrium, but also in the nonequilibrium state induced when the liquid is
subjected to a heat flux induced by a stationary thermal gradient. We also
evaluate the effect of the thermal gradient on the dynamic structure factor
of the fluid. However, since to our knowledge there are no experimental
results available in the literature for light scattering from a nematic in
this steady states, to estimate the predictions of our model we used
experimental parameter values similar to those used in light scattering
experiments for an isotropic simple fluid. In this way the effect of this
long-range behavior on the dynamic structure factor of the fluid is examined
for a geometry used in light scattering experiments. We find that the
external gradient introduces an asymmetry of the spectrum shifting its
maximum towards negative frequency intervals by an amount proportional to
the magnitude of the gradient which may be of the order of $\sim 7\%$. Also,
the width at half height may decrease by a factor of $\sim 10\%$.

\section{Model and Basic Equations}

Consider a thermotropic nematic liquid crystal layer of thickness $d$
confined between two parallel plates maintained at the uniform temperatures $%
T_{1}$ and $T_{2}$, respectively, in a homeotropic arrangement as depicted
in Fig. 1. The transverse dimensions of the cell along the $x$ and $y$
directions are large compared to $d$. The hydrodynamic state of the nematic
is specified by the velocity field, $\overrightarrow{v}\left( 
\overrightarrow{r},t\right) $, the unit vector defining the local symmetry
axis (director field), $\hat{n}\left( \overrightarrow{r},t\right) $, the
pressure, $p\left( \overrightarrow{r},t\right) $ and the temperature, $%
T\left( \overrightarrow{r},t\right) $. We assume that the nematic has
reached a stationary state characterized by $T_{ss}$, $p_{ss}$ and $\hat{n}%
_{ss}=\hat{e}_{z}$, without convection, e.g., $\overrightarrow{v}_{ss}=0$.
By symmetry $p_{ss}$ and $T_{ss}$ may only depend on $z$ and if we neglect
the variation of thermal conductivity with $z$,\ the stationary state is
defined by the solution of the equations for $p_{ss}$, $\rho _{ss}$ and $%
T_{ss}$, which are obtained from the general nematodynamic equations for
this geometry, see Appendix in Ref. \cite{camacho}. As a result, a linear
stationary temperature profile is established in the cell, $T_{ss}\left(
z\right) =T_{0}\left( 1+\overline{B}z\right) $, where $T_{0}=\left(
T_{1}+T_{2}\right) /2$ and $\overline{B}=\beta /d$ with $\beta =d\left(
dT_{ss}/dz\right) /T_{0}$.
\begin{figure}
\includegraphics{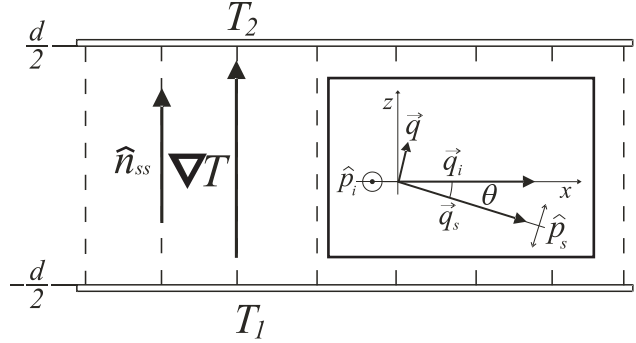}
\caption{Schematic representation of a plane homeotropic cell with
a constant thermal gradient along $z$ direction. The inset shows the
scattering geometry.The scattering angle is $\theta$.}
\end{figure}

We want to describe the dynamics of the spontaneous thermal (hydrodynamic)
fluctuations, $\delta p\left( \overrightarrow{r},t\right) =p\left( 
\overrightarrow{r},t\right) -p_{ss}$, etc, around this stationary state. A
complete set of stochastic equations for the space-time evolution of these
fluctuations is obtained by linearizing the general hydrodynamic equations 
\cite{buka}, \cite{landau2}, \cite{camacho} and by using the fluctuating
hydrodynamics formalism of Landau and Lifshitz \cite{landau1}. In this work
we will only calculate correlation functions in the bulk of the nematic,
that is, for points $\overrightarrow{r}$ and $\overrightarrow{r}^{\prime }$
with $z$ and $z^{\prime }$ far away from the boundaries, which means, $%
\left\vert z-z^{\prime }\right\vert \ll d$ and $\left\vert z+z^{\prime
}\right\vert \ll d$. Moreover, since the cartesian components of the
director and velocity fields turn out to be strongly coupled, it will be
convenient to work in a representation where this coupling is minimum. This
is accomplished by defining the transverse component of the director
fluctuating field as its projection along the perpendicular direction to the 
$\overrightarrow{k}-\hat{n}_{ss}$ plane, i.e., $\delta \tilde{n}%
_{1}=k_{\perp }^{-1}\hat{n}_{ss}\cdot \left( \overrightarrow{k}\times \delta 
\widetilde{\overrightarrow{n}}\right) $, where $k_{\perp }=\left(
k_{x}^{2}+k_{y}^{2}\right) ^{1/2}$ and the upper tilde denotes the
space-Fourier transform of fields. Similarly, the longitudinal component, $%
\delta \tilde{n}_{3}=k^{-1}\overrightarrow{k}\cdot \delta \widetilde{%
\overrightarrow{n}}$, is its projection in the $k-\hat{n}_{ss}$ plane.

In previous work we have shown that for a thermotropic nematic the modes
associated with the director relaxation are much slower than the visco-heat
and sound modes \cite{camacho}. For typical material parameter values of a
thermotropic nematic the ratio of these relaxation times is $\tau
_{orientation}/\tau _{visco-heat}\sim 10^{5}$. Therefore, transverse and
longitudinal director fluctuations relax to equilibrium much more slowly
than fluctuations in temperature, pressure and velocity fields. Actually,
the existence of widely separated time-scales may be exploited to eliminate
the fast variables from the general dynamic equations obtaining a reduced
description in which only the slow variables are involved. The time scaling
perturbation method introduced by Geigenm\"{u}ller et al. \cite{geigen1}, 
\cite{geigen3} may be implemented in order\ to diminish the couplings
between nematodynamic fluctuations. It allows us to find, on the slow
time-scales, a contracted description in terms of the slow variables only
with a reduced dynamic matrix which can be constructed by a perturbation
procedure.

By using this method it can be shown that in the slow time-scale, director
fluctuations $\delta \tilde{n}_{1}$ and $\delta \tilde{n}_{3}$ obey the
stochastic equations \cite{hijar2},%
\begin{equation}
\partial _{t}\delta \tilde{n}_{\mu }=-\omega _{\mu }\left( \overrightarrow{k}%
\right) \delta \tilde{n}_{\mu }-\tilde{\sigma}_{\mu },\text{ }\mu =\left\{ 1%
\text{ transverse, }3\text{ longitudinal }\right.  \label{10}
\end{equation}%
where we have introduced the following abbreviations%
\begin{equation}
\omega _{1}\left( \overrightarrow{k}\right) =\frac{1}{\gamma _{1}}\left(
K_{2}k_{\perp }^{2}+K_{3}k_{z}^{2}\right) \left[ 1+\frac{1}{4}\frac{\gamma
_{1}\left( 1+\lambda \right) ^{2}k_{z}^{2}}{\nu _{2}k_{\perp }^{2}+\nu
_{3}k_{z}^{2}}\right] ,  \label{11}
\end{equation}%
\begin{equation}
\omega _{3}\left( \vec{k}\right) =\frac{1}{\gamma _{1}}\left( K_{1}k_{\perp
}^{2}+K_{3}k_{z}^{2}\right) \left\{ 1+\frac{1}{4}\frac{\gamma _{1}\left[
\left( 1+\lambda \right) k_{z}^{2}+\left( 1-\lambda \right) k_{\perp }^{2}%
\right] ^{2}}{\nu _{3}k_{\perp }^{4}+2\left( \nu _{1}+\nu _{2}-\nu
_{3}\right) k_{\perp }^{2}k_{z}^{2}+\nu _{3}k_{z}^{4}}\right\} .  \label{12}
\end{equation}%
The fluctuating sources $\tilde{\sigma}_{\mu }$ are defined as 
\begin{equation}
\tilde{\sigma}_{1}=\frac{1}{k_{\perp }}\left[ k_{x}\widetilde{\Upsilon }%
_{y}-k_{y}\widetilde{\Upsilon }_{x}+\frac{1}{2}\frac{\left( 1+\lambda
\right) k_{z}}{\nu _{2}k_{\perp }^{2}+\nu _{3}k_{z}^{2}}\left( k_{x}k_{j}%
\widetilde{\Sigma }_{yj}-k_{y}k_{j}\widetilde{\Sigma }_{yj}\right) \right] ,
\label{13}
\end{equation}%
\begin{equation}
\tilde{\sigma}_{3}=\frac{1}{k}\left[ k_{x}\widetilde{\Upsilon }_{x}+k_{y}%
\widetilde{\Upsilon }_{y}+\frac{1}{2}\frac{\left( 1+\lambda \right)
k_{z}^{2}+\left( 1-\lambda \right) k_{\perp }^{2}}{\nu _{3}k_{\perp
}^{4}+2\left( \nu _{1}+\nu _{2}-\nu _{3}\right) k_{\perp }^{2}k_{z}^{2}+\nu
_{3}k_{z}^{4}}\left( k^{2}k_{j}\widetilde{\Sigma }_{zj}-k_{z}k_{i}k_{j}%
\widetilde{\Sigma }_{ij}\right) \right] ,  \label{14}
\end{equation}%
where $\lambda =\gamma _{1}/\gamma _{2}$\ is a non dissipative coefficient
associated with director relaxation, $\gamma _{1}$\ and $\gamma _{2}$\ being
orientational viscosities; $K_{1}$, $K_{2}$\ and $K_{3}$\ are the splay,
twist and bend elastic constants, respectively; and $\nu _{1}$, $\nu _{2}$\
and $\nu _{3}$\ are three shear viscosity coefficients. The stochastic
components of the quasi-current of director field and the stress tensor, $%
\Upsilon _{i}$ and $\Sigma _{ij}$, respectively, are zero averaged $%
\left\langle \Upsilon _{i}\right\rangle =\left\langle \Sigma
_{ij}\right\rangle =0$ and in equilibrium they satisfy the fluctuation
dissipation relations \cite{camacho}%
\begin{equation}
\left\langle \Upsilon _{i}\left( \overrightarrow{r},t\right) \Upsilon
_{j}\left( \overrightarrow{r}^{\prime },t^{\prime }\right) \right\rangle
=2k_{B}T\frac{1}{\gamma _{1}}\delta _{ij}^{\perp }\delta \left( 
\overrightarrow{r}-\overrightarrow{r}^{\prime }\right) \delta \left(
t-t^{\prime }\right) ,  \label{16}
\end{equation}%
\begin{equation}
\left\langle \Sigma _{ij}\left( \overrightarrow{r},t\right) \Sigma
_{lm}\left( \overrightarrow{r}^{\prime },t^{\prime }\right) \right\rangle
=2k_{B}T\nu _{ijlm}\delta \left( \overrightarrow{r}-\overrightarrow{r}%
^{\prime }\right) \delta \left( t-t^{\prime }\right) ,  \label{17}
\end{equation}%
where $k_{B}$ is Boltzmann's constant, $T$ is the equilibrium temperature of
the nematic, $\delta _{ij}^{\perp }=\delta _{ij}-n_{i}n_{s}$ is a projection
operator and $\nu _{ijlm}$ is the viscous tensor as given in Ref. \cite%
{camacho}. Relations (\ref{16}) and (\ref{17}) describe stationary Gaussian
Markov processes.

Now, we will assume that in the non-equilibrium steady state the fluctuation
dissipation relations for $\Upsilon _{i}$ and $\Sigma _{ij}$ can be obtained
from (\ref{16}) and (\ref{17}) by replacing the equilibrium temperature $T$
by the local $z$-dependent stationary temperature $T_{ss}\left( z\right) $.
This assumption leads to%
\begin{equation}
\left\langle \hat{\Upsilon}_{i}\left( \overrightarrow{k},\omega \right) \hat{%
\Upsilon}_{j}\left( \overrightarrow{k}^{\prime },\omega ^{\prime }\right)
\right\rangle _{ss}=2\left( 2\pi \right) ^{4}k_{B}T_{0}\frac{1}{\gamma _{1}}%
\delta _{ij}^{\perp }\left( 1+i\overline{B}\frac{\partial }{\partial k_{z}}%
\right) \delta \left( \overrightarrow{k}+\overrightarrow{k}^{\prime }\right)
\delta \left( \omega +\omega ^{\prime }\right) ,  \label{18}
\end{equation}%
\begin{equation}
\left\langle \hat{\Sigma}_{ij}\left( \overrightarrow{k},\omega \right) \hat{%
\Sigma}_{lm}\left( \overrightarrow{k}^{\prime },\omega ^{\prime }\right)
\right\rangle _{ss}=2\left( 2\pi \right) ^{4}k_{B}T_{0}\nu _{ijlm}\left( 1+i%
\overline{B}\frac{\partial }{\partial k_{z}}\right) \delta \left( 
\overrightarrow{k}+\overrightarrow{k}^{\prime }\right) \delta \left( \omega
+\omega ^{\prime }\right) .  \label{19}
\end{equation}

\section{Orientational Correlation Functions}

The space-time Fourier transform of Eq. (\ref{10}) is $\delta \hat{n}_{\mu
}\left( \vec{k},\omega \right) =\left[ -i\omega +\omega _{\mu }\left( \vec{k}%
\right) \right] ^{-1}\hat{\sigma}_{\mu }\left( \vec{k},\omega \right) $ and
the director auto-correlation functions $\hat{X}_{\mu \mu }\left( 
\overrightarrow{k},\overrightarrow{k}^{\prime };\omega ,\omega ^{\prime
}\right) =\left\langle \delta \hat{n}_{\mu }\left( \overrightarrow{k},\omega
\right) \delta \hat{n}_{\mu }\left( \overrightarrow{k}^{\prime },\omega
^{\prime }\right) \right\rangle _{ss}$can be constructed in terms of the
corresponding fluctuation-dissipation relations for $\hat{\sigma}_{1}$ and $%
\hat{\sigma}_{3}$,%
\begin{equation}
\hat{X}_{\mu \mu }=\frac{\left\langle \hat{\sigma}_{\mu }\left( 
\overrightarrow{k},\omega \right) \hat{\sigma}_{\mu }\left( \overrightarrow{k%
}^{\prime },\omega ^{\prime }\right) \right\rangle _{ss}}{\left[ -i\omega
+\omega _{\mu }\left( \overrightarrow{k}\right) \right] \left[ -i\omega
^{\prime }+\omega _{\mu }\left( \overrightarrow{k}^{\prime }\right) \right] }%
,  \label{21}
\end{equation}%
where no summation over the repeated index $\mu $ is implied. In these
equations $\hat{f}$ denotes the space-time Fourier transform of the field $f$%
. The fluctuation-disspiation relations obeyed by $\hat{\sigma}_{1}$ and $%
\hat{\sigma}_{3}$ may be found from Eqs. (\ref{18}) and (\ref{19}). By
inserting the resulting expressions into Eq. (\ref{21}) we obtain explicit
forms for $\hat{X}_{\mu \mu }$. First we shall examine the spatial limiting
behavior\ of these correlations which is given by 
\begin{equation}
X_{\mu \mu }\left( \overrightarrow{r},\overrightarrow{r}^{\prime
};t,t^{\prime }\right) =\frac{1}{\left( 2\pi \right) ^{8}}\int d%
\overrightarrow{k}d\overrightarrow{k}^{\prime }d\omega d\omega ^{\prime
}e^{i\left( \overrightarrow{k}\cdot \vec{r}+\overrightarrow{k}^{\prime
}\cdot \overrightarrow{r}^{\prime }-\omega t-\omega ^{\prime }t^{\prime
}\right) }\hat{X}_{\mu \mu }\left( \overrightarrow{k},\overrightarrow{k}%
^{\prime };\omega ,\omega ^{\prime }\right) ,  \label{22}
\end{equation}%
when $(t^{\prime }-t)\longrightarrow 0$. Since the calculation of $X_{11}$
and $X_{33}$ is formally the same, we will only describe the procedure for $%
X_{11}$. Calculating $\left\langle \hat{\sigma}_{1}\left( \overrightarrow{k}%
,\omega \right) \hat{\sigma}_{1}\left( \overrightarrow{k}^{\prime },\omega
^{\prime }\right) \right\rangle _{ss}$ with the help of Eqs. (\ref{13}), (%
\ref{18}) and (\ref{19}), replacing the result into (\ref{21}) and
integrating over $\omega ^{\prime }$, $\overrightarrow{k}^{\prime }$ and $%
\omega $, we find that%
\begin{equation}
X_{11}\left( \overrightarrow{r},\overrightarrow{r}^{\prime };t,t^{\prime
}\right) =X_{11}^{(1)}\left( \overrightarrow{r},\overrightarrow{r}^{\prime
};t,t^{\prime }\right) +X_{11}^{(2)}\left( \overrightarrow{r},%
\overrightarrow{r}^{\prime };t,t^{\prime }\right) .  \label{23}
\end{equation}%
where%
\begin{equation}
X_{11}^{(1)}=-\frac{k_{B}T_{ss}\left( z\right) }{\left( 2\pi \right) ^{3}}%
\int d\vec{k}\frac{e^{i\vec{k}\cdot \left( \overrightarrow{r}-%
\overrightarrow{r}^{\prime }\right) -\left\vert t-t^{\prime }\right\vert
\omega _{1}\left( \vec{k}\right) }}{K_{2}k_{\perp }^{2}+K_{3}k_{z}^{2}}
\label{24}
\end{equation}%
and 
\begin{equation}
X_{11}^{(2)}=-\frac{ik_{B}T_{0}\overline{B}}{\left( 2\pi \right) ^{3}\gamma
_{1}}\int d\vec{k}k_{z}\left[ -\frac{\gamma _{1}K_{3}}{\left( K_{1}k_{\perp
}^{2}+K_{3}k_{z}^{2}\right) ^{2}}-2\left\vert t-t^{\prime }\right\vert
b_{1}\left( \vec{k}\right) \right] e^{i\vec{k}\cdot \left( \overrightarrow{r}%
-\overrightarrow{r}^{\prime }\right) -\left\vert t-t^{\prime }\right\vert
\omega _{1}\left( \vec{k}\right) }.  \label{25}
\end{equation}%
Here we have considered $t^{\prime }>t$ and defined%
\begin{equation}
b_{1}\left( \vec{k}\right) =\left( \frac{1+\lambda }{2}\right) ^{2}\frac{%
\gamma _{1}\nu _{2}k_{\perp }^{2}}{\left( \nu _{2}k_{\perp }^{2}+\nu
_{3}k_{z}^{2}\right) ^{2}}+\frac{K_{3}}{K_{2}k_{\perp }^{2}+K_{3}k_{z}^{2}}%
\left[ 1+\frac{1}{4}\frac{\gamma _{1}\left( 1+\lambda \right) ^{2}k_{z}^{2}}{%
\nu _{2}k_{\perp }^{2}+\nu _{3}k_{z}^{2}}\right] \text{.}  \label{26}
\end{equation}

From Eqs. (\ref{24}) and (\ref{25}) we can find the spatial range order of
transverse director fluctuations by considering the limiting behavior at
small times $\left\vert t-t^{\prime }\right\vert $ or large distances $%
\left\vert \overrightarrow{r}-\overrightarrow{r}^{\prime }\right\vert $,
that is, when $\xi =K\left\vert t-t^{\prime }\right\vert /\nu \left\vert 
\overrightarrow{r}-\overrightarrow{r}^{\prime }\right\vert ^{2}\ll 1$, which
corresponds to the limit of static correlation functions. In this limit we
find%
\begin{equation}
\lim_{\xi \rightarrow 0}\left\vert X_{11}\right\vert =\frac{%
k_{B}T_{ss}\left( z\right) }{4\pi \left( K_{2}K_{3}\right) ^{1/2}}\frac{1}{%
\left[ \left\vert \overrightarrow{r}_{\perp }-\overrightarrow{r}_{\perp
}^{\prime }\right\vert ^{2}+\frac{K_{2}}{K_{3}}\left( z-z^{\prime }\right)
^{2}\right] ^{1/2}}\left[ 1-\left( \frac{dT_{ss}}{dz}\right) \frac{%
z-z^{\prime }}{2T_{ss}\left( z\right) }\right] ,  \label{28}
\end{equation}%
where $\left\vert \overrightarrow{r}_{\perp }-\overrightarrow{r}_{\perp
}^{\prime }\right\vert ^{2}=\left( x-x^{\prime }\right) ^{2}+\left(
y-y^{\prime }\right) ^{2}$.

Similarly, for the longitudinal director fluctuations we obtain%
\begin{eqnarray}
\lim_{\xi \rightarrow 0}\left\vert X_{33}\right\vert  &=&\frac{%
k_{B}T_{ss}\left( z\right) }{4\pi \left( K_{1}-K_{3}\right) }\left\{ \frac{1%
}{\left\vert \overrightarrow{r}-\overrightarrow{r}^{\prime }\right\vert }-%
\frac{\left( K_{3}/K_{1}\right) ^{1/2}}{\left[ \left\vert \overrightarrow{r}%
_{\perp }-\overrightarrow{r}_{\perp }^{\prime }\right\vert ^{2}+\frac{K_{1}}{%
K_{3}}\left( z-z^{\prime }\right) ^{2}\right] ^{1/2}}\right\}   \notag \\
&&\times \left[ 1-\left( \frac{dT_{ss}}{dz}\right) \frac{z-z^{\prime }}{%
2T_{ss}\left( z\right) }\right] ,  \label{29}
\end{eqnarray}%
Note that in equilibrium, $\overline{B}=0$, $T_{ss}\left( z\right) =T$, a
constant, and the transverse orientational correlation functions reduces to%
\begin{equation}
\lim_{\xi \rightarrow 0}\left\vert X_{11}^{eq}\right\vert =\frac{k_{B}T}{%
4\pi \left( K_{2}K_{3}\right) ^{1/2}}\frac{1}{\left[ \left\vert 
\overrightarrow{r}_{\perp }-\overrightarrow{r}_{\perp }^{\prime }\right\vert
^{2}+\frac{K_{2}}{K_{3}}\left( z-z^{\prime }\right) ^{2}\right] ^{1/2}},
\label{30}
\end{equation}%
while the corresponding correlation function for the longitudinal director
components is 
\begin{equation}
\lim_{\xi \rightarrow 0}\left\vert X_{33}^{eq}\right\vert =\frac{k_{B}T}{%
4\pi \left( K_{1}-K_{3}\right) }\left\{ \frac{1}{\left\vert \overrightarrow{r%
}-\overrightarrow{r}^{\prime }\right\vert }-\frac{\left( K_{3}/K_{1}\right)
^{1/2}}{\left[ \left\vert \overrightarrow{r}_{\perp }-\overrightarrow{r}%
_{\perp }^{\prime }\right\vert ^{2}+\frac{K_{1}}{K_{3}}\left( z-z^{\prime
}\right) ^{2}\right] ^{1/2}}\right\} .  \label{31}
\end{equation}

Note that both $X_{11}^{eq}$ and $X_{33}^{eq}$ are long ranged as could have
been anticipated due to the well known property of a nematic which
spontaneously exhibits a macroscopic orientational order. They decay
algebraically as $\left\vert \overrightarrow{r}-\overrightarrow{r}^{\prime
}\right\vert ^{-1}$. In Fig. 2 we plot the normalized static correlation in
equilibrium $\overline{X}_{11}^{eq}\equiv 4\pi d\left( K_{2}K_{3}\right)
^{1/2}X_{11}^{eq}/k_{B}T_{0}$ for $x-x^{\prime }=y-y^{\prime }=z=0$, $%
T=T_{0} $ and as a function of normalized distance $z^{\prime }/d$. In the
stationary state, $\overline{B}\neq 0$, \ the behavior of both correlations
is modified by the presence of a term proportional to the temperature
gradient which behaves as $\left\vert \overrightarrow{r}-\overrightarrow{r}%
^{\prime }\right\vert ^{0}$, that is, which does not decay, in the direction
of the temperature gradient. This is also depicted in Fig. 2, where we plot
the normalized static correlation $\overline{X}_{11}\equiv 4\pi d\left(
K_{2}K_{3}\right) ^{1/2}X_{11}/k_{B}T_{0}$ for the same conditions as in the
equiluibrium case and a value of the normalized thermal gradient $\beta
^{\prime }=dTL_{0}^{-1}\left( dT_{ss}/dz\right) =0.5>0$\emph{.} \ Note that
the because of the presence of $dT_{ss}/dz$, $\overline{X}_{11}$ becomes
asymmetric. When compared to its equilibrium value, correlations increase in
the direction of lower temperatures and decrease in the opposite direction.
Indeed, for a plate separation $d=10^{-2}$ cm, the difference between both
curves becomes significant for $z^{\prime }/d\sim 10^{-1}$. This means that
the wavevectors sensitive to this difference are of the order of $q\sim
10^{3}$ cm$^{-1}$. Since for light scattering the wavevector $k$ and the
scattering angle $\theta $ are related by $q=2q_{i}\sin \theta /2$, where $%
q_{i}\sim 10^{5}$cm$^{-1}$\ is the incident wave number, this implies very
low scattering angles $\theta \sim 0.1^{\circ }$. A quantitative evaluation
of this nonequilibrium effect on a measurable property will be discussed in
the next section for the structure factor of the fluid.

\begin{figure}
\includegraphics{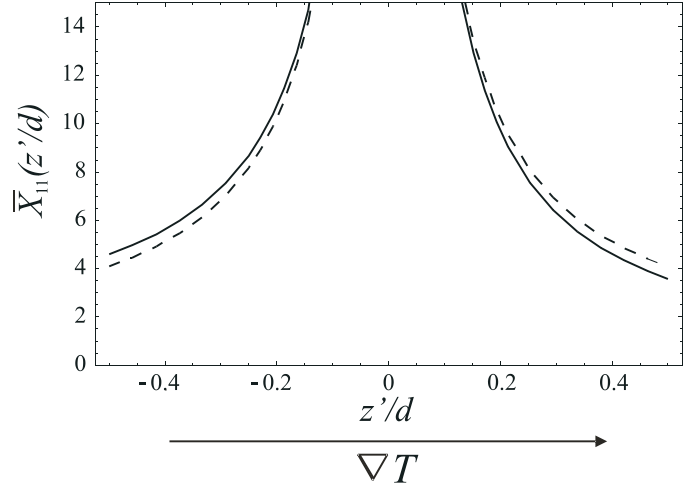}
\caption{(- - -) Decay of $\overline{X}_{11}^{eq}$ as a function of $%
z^{\prime }/d$. (-----) Decay of $\overline{X}_{11}$
as function of $z^{\prime} /d$ for $\beta ^{\prime
}=0.5$. The elastic constants values are $K_{2}=4.4 \times 10^{-7}$ dyn , $%
K_{3}=8.9 \times 10^{-7}$ dyn.}
\end{figure}

\section{Light Scattering Spectrum}

As an application of the previous theory we now calculate the light
scattering spectrum of the nematic in the stationary state and for the
scattering geometry defined in Fig. 1. For a nematic dielectric tensor
fluctuations come mainly from director fluctuations and for the present
model the spectral intensity of the scattered light is proportional to the
dynamic structure factor which is given by the real part of $\left\langle
\delta \hat{n}_{1}\left( \overrightarrow{q},\omega \right) \delta \hat{n}%
_{1}\left( -\overrightarrow{q},-\omega \right) \right\rangle $, $C\left( 
\overrightarrow{q},\omega \right) $, 
\begin{equation}
S\left( \overrightarrow{q},\omega \right) =-\frac{\varepsilon _{a}^{2}\cos
^{2}\theta }{V_{s}t_{s}}C\left( \overrightarrow{q},\omega \right) ,
\label{34}
\end{equation}%
where $V_{s}$ and $t_{s}$ are the scattering volume and scattering time,
respectively, $\overrightarrow{q}=\overrightarrow{q}_{i}-\overrightarrow{q}%
_{s}$ is the scattering vector ($q_{y}=0$) and $\omega =\omega _{s}-\omega
_{i}$ is the frequency shift. $\varepsilon _{a}=\varepsilon _{\parallel
}-\varepsilon _{\perp }$ denotes the dielectric constant anisotropy.
Evaluating this expression with the help of Eqs. (\ref{18}), (\ref{19}) we
obtain 
\begin{equation}
S\left( \overrightarrow{q},\omega \right) =\frac{2\varepsilon
_{a}^{2}k_{B}T_{0}\cos ^{2}\theta }{\gamma _{1}}\frac{\alpha \left( 
\overrightarrow{q}\right) }{\omega ^{2}+\omega _{1}^{2}\left( 
\overrightarrow{q}\right) }\left\{ 1-\frac{1}{T_{0}}\left( \frac{dT_{ss}}{dz}%
\right) \frac{2\omega q_{z}\beta \left( \overrightarrow{q}\right) }{\omega
^{2}+\omega _{1}^{2}\left( \overrightarrow{q}\right) }\right\} .  \label{36}
\end{equation}%
In this result we have considered the spectrum produced by a scattering
volume located at the center of the cell. The nonequilibrium contribution
has been written up to the smallest power of the wave number $q$, that is,
up to the leading term in the hydrodynamic limit $q\rightarrow 0$, $\omega
_{1}\left( \overrightarrow{q}\right) $ is given by Eq. (\ref{11}) and the
functions $\alpha \left( \overrightarrow{q}\right) $ and $\beta \left( 
\overrightarrow{q}\right) $ contain angular information through 
\begin{equation}
\alpha \left( \overrightarrow{q}\right) =1+\left( \frac{1+\lambda }{2}%
\right) ^{2}\frac{\gamma _{1}q_{z}^{2}}{\nu _{2}q_{x}^{2}+\nu _{3}q_{z}^{2}}
\label{37}
\end{equation}%
and%
\begin{equation}
\beta \left( \overrightarrow{q}\right) =\frac{1}{\gamma _{1}}\left[
K_{3}\alpha \left( \overrightarrow{q}\right) +\left(
K_{2}q_{x}^{2}+K_{3}q_{z}^{2}\right) \frac{\gamma _{1}\nu _{2}q_{x}^{2}}{%
\left( \nu _{2}q_{x}^{2}+\nu _{3}q_{z}^{2}\right) ^{2}}\right] .  \label{38}
\end{equation}

In equilibrium, (\ref{36}) reduces to%
\begin{equation}
S^{eq}\left( \overrightarrow{q},\omega \right) =\frac{2\varepsilon
_{a}^{2}k_{B}T\cos ^{2}\theta }{\gamma _{1}}\frac{\alpha \left( 
\overrightarrow{q}\right) }{\omega ^{2}+\omega _{1}^{2}\left( 
\overrightarrow{q}\right) },  \label{39}
\end{equation}%
which behaves as $q^{-4}$. This dependence is responsible for the well known
long-range order spatial behavior of the orientational correlations
exhibited spontaneously by a nematic in equilibrium. On the other hand, the
nonequilibrium contribution is%
\begin{equation}
S^{neq}\left( \overrightarrow{q},\omega \right) =-\frac{2\varepsilon
_{a}^{2}k_{B}\sin ^{2}\theta }{\gamma _{1}}\left( \frac{dT_{ss}}{dz}\right) 
\frac{2\omega q_{z}\alpha \left( \overrightarrow{q}\right) \beta \left( 
\overrightarrow{q}\right) }{\left[ \omega ^{2}+\omega _{1}^{2}\left( 
\overrightarrow{q}\right) \right] ^{2}},  \label{40}
\end{equation}%
which also shows a long range-order decaying as $q^{-5}$. Note that close to
equilibrium the size of the shift is indeed proportional to $dT_{ss}/dz$.
Its odd dependence on $\omega $ introduces an asymmetry in the shape of the
structure factor, shifting the maximum towards the region of negative values
of $\omega $, as shown in Fig. 3.

For a fixed scattering vector $\overrightarrow{q}$, we define the
dimensionless dynamic structure factor, $S_{0}\left( \omega _{0}\right) $,
in terms of the normalized thermal gradient $\beta ^{\prime
}=dT_{0}^{-1}\left( dT_{ss}/dz\right) $ and the normalized frequency $\omega
_{0}=\omega /\omega _{1}\left( \overrightarrow{q}\right) $, by%
\begin{equation}
S_{0}\left( \omega _{0}\right) =\frac{S\left( \overrightarrow{q},\omega
\right) }{S^{eq}\left( \overrightarrow{q},0\right) }=\frac{1}{1+\omega
_{0}^{2}}\left\{ 1-\beta ^{\prime }\frac{2A\omega _{0}}{1+\omega _{0}^{2}}%
\right\}  \label{41}
\end{equation}%
where $A=q_{z}\beta \left( \overrightarrow{q}\right) /\omega _{1}\left( 
\overrightarrow{q}\right) d$. In Fig. 3 we compare $S_{0}\left( \omega
_{0}\right) $ and $S_{0}^{eq}\left( \omega _{0}\right) $ for low scattering
angles, $\theta \sim 0.1%
{{}^\circ}%
$, $k_{i}\sim 10^{5}$ cm$^{-1}$, $\beta ^{\prime }=0.5$ and typical values
of material parameters of a thermotropic nematic \cite{khoo}. From this
result it follows that the increase of the maximum is about 7\% and that the
decrease in the half width at half height is 10\%. This shows that the
nonequilibrium state may induce changes in the dynamic structure factor
which might be detected experimentally.
\begin{figure}
\includegraphics{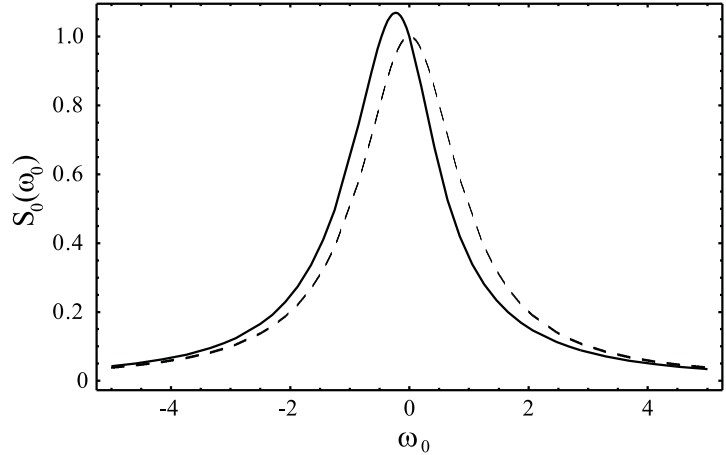}
\caption{(- - -) Normalized structure factor at equilibrium
$S_{0}^{eq}\left( \omega _{0}\right) $. (-----) Nonequilibrium structure
factor $S_{0}\left( \omega_{0}\right) $. We use $\beta ^{\prime }=0.5$,
$d=10^{-1}cm$, $\theta \sim 0.1%
{{}^\circ}%
$, $k_{i}\sim 10^{5}$ $cm^{-1}$. The values of the elastic constants are the
same as in Fig. 2, $\nu _{2}=0.41$ poise, $\nu _{3}=0.24$ poise, $\gamma
_{1}=1.03$ poise, $\lambda =1.03$.}
\end{figure}

\section{Discussion}

We have shown theoretically that the orientation correlation functions for a
nematic liquid crystal exhibit long-range order both, in equilibrium and for
a steday state induced by an external thermal gradient. We also estimated
the influence of the stationary heat flux on the light scattering spectrum
of a thermotropic nematic. The analysis carried out in this work included
only the orientation correlation functions and the model has been
constructed so that it corresponds to an experimental arrangement
appropiated to detect the so called mode 2 of the spectrum \cite{handbook}.
However, it should be emphasized that this is a model calculation and since
to our knowledge there are no experimental results to compare with, the
correctness of our choice of experimental parameters such as $d$, $\theta $
or $\beta ^{\prime }$, remains to be assessed. However, as found in other
nonequilibrium states for liquid crystals \cite{camacho}, their magnitude
suggests that they might be experimentally detected.

It is worth pointing out that in previous work we have analyzed the effects
produced by a nonequilibrium state induced by a concentration gradient in
Ref. \cite{hijar3}. Other correlation functions such as velocity-velocity or
temperature-temperature may also be calculated by using the same approach.
These correlations turn out to be also long-ranged in the presence of the
thermal gradient as reported in Ref. \cite{hijar2}, \cite{hijar1}. They are
also used to compare the physical mechanisms responsible for the appearing
of long-range order in stationary states of nematics.

\begin{acknowledgments}
We acknowledge partial financial support from DGAPA-UNAM IN112503 and from
FENOMEC through grant CONACYT 400316-5-G25427E, M\'{e}xico.
\end{acknowledgments}

\end{document}